\documentclass[]{spie}  %>>> use for US letter paper
\usepackage{amsmath,amsfonts,amssymb}
\usepackage{graphicx}
\usepackage{algorithm2e}
\usepackage[colorlinks=true, allcolors=blue]{hyperref}

\title{Self-optimizing adaptive optics control with Reinforcement Learning}

\author[a,*]{R. Landman}
\author[a]{S.Y. Haffert}
\author[a]{V.M. Radhakrishnan}
\author[a]{C.U. Keller}

\affil[a]{Leiden Observatory, Leiden University, PO Box 9513, 2300 RA Leiden, The Netherlands}

\authorinfo{*R.Landman: E-mail: rlandman@strw.leidenuniv.nl}

\begin{document} 
\maketitle

\begin{abstract}
Current and future high-contrast imaging instruments require extreme Adaptive Optics (XAO) systems to reach contrasts necessary to directly image exoplanets. Telescope vibrations and the temporal error induced by the latency of the control loop limit the performance of these systems. Optimization of the (predictive) control algorithm is crucial in reducing these effects. We describe how model-free Reinforcement Learning can be used to optimize a Recurrent Neural Network controller for closed-loop adaptive optics control. We verify our proposed approach for tip-tilt control in simulations and a lab setup. The results show that this algorithm can effectively learn to suppress a combination of tip-tilt vibrations. Furthermore, we report decreased residuals for power-law input turbulence compared to an optimal gain integrator. Finally, we demonstrate that the controller can learn to identify the parameters of a varying vibration without requiring online updating of the control law. We conclude that Reinforcement Learning is a promising approach towards data-driven predictive control; future research will apply this approach to the control of high-order deformable mirrors.
\end{abstract}

% Include a list of keywords after the abstract 

\section{INTRODUCTION}
    One of the main limitations of current and future ground-based high contrast imaging (HCI) instruments is the wavefront error induced by the time lag between sensing and correcting the wavefront. This time lag results in a halo of speckles along the wind direction, which limits the contrast at small separations \cite{Cantalloube2020_WDH_sphere}. Furthermore, the finite control bandwidth results in low-order residuals after the correction. One of the main causes of these residuals are vibrations close to the maximum control frequency that originate from a variety of sources. All ground-based high-contrast imaging instruments suffer from vibrations including SCExAO \cite{Lozi2018_vibration_scexao}, GPI \cite{Hartung2014_vibration_GPI} and SPHERE \cite{Sauvage2010_saxo}. These vibrations decrease the resolution of long-exposure images. Furthermore, some coronagraphs are highly sensitive to residual tip-tilt jitter\cite{Lloyd2005_Lyot_tiptilt, Mawet2010_vector_vortex}, which will further reduce the contrast.
    
    Both of these effects may be reduced by the use of better control algorithms. Early work on improved controllers involved optimizing the modal gains of a modal integrator based on the open-loop temporal Power Spectral Density (PSD)\cite{Gendron1995_OMGI_I,Gendron1995_OMGI_II}. This optimal gain is a trade-off between turbulence rejection and the amplification of noise and disturbances outside the control bandwidth. However, to reduce the effect of the servo-lag we have to predict the disturbances before they are measured. A popular approach to optimal control is Linear Quadratic Gaussian (LQG) control\cite{Paschall1993_LQG}, which is based on a Kalman filter. Variants of LQG-based controllers have been demonstrated to improve the performance in tip-tilt vibration mitigation in simulations \cite{Correia2012_vibrations}, lab verifications \cite{Petit2008_LQG_lab_vibration} and on-sky\cite{Sivo2014_LQG_canary, Petit2014_SPHERE_onsky, Hartung2014_vibration_GPI}. A major disadvantage of LQG-like controllers is their reliance on a state-space model of the turbulence and deformable mirror (DM) dynamics\cite{Correia2010_DM_dynamics}; this requires an accurate calibration of the model parameters. Hence, these controllers can also not correct for correlations in the turbulence that are not included in the model. A combination of an LQG controller for low-order vibrations and an optimal modal gain integrator for high order modes is currently being used in the AO systems of both SPHERE \cite{Petit2014_SPHERE_onsky} and GPI \cite{Poyneer2016_GPI}.
    
    Another group of data-driven algorithms construct a linear filter that minimizes the residual phase variance from (pseudo) open-loop data. This idea was originally proposed and demonstrated for modal control in \citenum{Dessenne1997}. More recently, a spatio-temporal filter based on Empirical Orthogonal Functions (EOFs) was demonstrated to significantly increase the contrast in a simulated HCI system\cite{Guyon2017_eof}. Performance of such linear filters was tested on open-loop AO telemetry for the Keck II AO system \cite{Jensen-Clem2019_EOF} and SPHERE\cite{VanKooten2020_PC} and showed a decrease in residual wavefront error. An issue with these data-driven predictive control algorithms is that they rely on the prediction of (pseudo) open-loop wavefront data, while XAO systems operate in closed-loop. To use these controllers, one has to reconstruct the pseudo open-loop residuals. Not only does this require knowledge of the servo-lag of the system and DM dynamics, it is also not trivial in the case where the response of the wavefront sensor is nonlinear, such as for the Pyramid Wavefront sensor. Furthermore, a relatively small error on the reconstruction or prediction of the pseudo open-loop wavefront may be detrimental to the closed-loop contrast as opposed to a relative error on the closed-loop wavefront. 
    
   Another issue for these approaches is non-stationary turbulence \cite{VanKooten2019_nonstationary}. All these methods require online updating of the control parameters based on the current turbulence parameters. This updating of the control law is computationally very expensive and often involves inverting large matrices\cite{Gray2014_ensemble_LQG_nonstationary, Guyon2017_eof}. Recently, it was shown that a single Artificial Neural Networks (ANN) can learn to predict the wavefront for varying wind speeds and directions for a simulated AO system\cite{Liu2020_LSTM}. This eliminates the need for updating the control law to keep up with the changing turbulence.
    
   Here, we propose to use Reinforcement Learning (RL) for the data-driven optimization of the closed-loop adaptive optics control law. Machine Learning has recently received more attention in the AO community for nonlinear wavefront reconstruction \cite{Landman2020, Norris2020_photonic_wfs} and wavefront prediction \cite{Liu2020_LSTM,Swanson2018_CNN_prediction}. Furthermore, Reinforcement Learning was suggested as a method to directly optimize the contrast in a high contrast imaging system \cite{Radhakrishnan2018} and for model-free focal-plane wavefront correction \cite{Sun2018_EM_FPWC}. Our approach has some distinct advantages:
    
    \begin{itemize}
        \item \textbf{Model-free}:  The algorithm is completely self-optimizing and data-driven. This means that it does not require a model of the disturbances or system dynamics, or knowledge of system parameters.
        
        \item \textbf{Flexible}: The algorithm allows for an arbitrary optimization objective. The user is free to choose the metric to optimize based on the science goals of the instrument. Furthermore, it can handle multiple inputs, such as in a system with multiple wavefront sensors.
        
        \item \textbf{Nonlinear}: The proposed method can optimize any parameterized control law. For example, one can optimize the parameters of a nonlinear Artificial Neural Network (ANN) controller. This allows the controller to model the nonlinear dynamics of the system, such as for systems with a Pyramid Wavefront Sensor.
        
        \end{itemize}
    This paper is structured as follows: Section \ref{sec:Methods} provides an introduction to Reinforcement Learning and describes the relevant algorithm. Section \ref{sec:tip_tilt_simulations} simulates the algorithm's performance in tip-tilt control for vibrations and a power-law input, while section \ref{sec:tip_tilt_lab} shows the results from a lab experiment. Finally, section \ref{sec:conclusions} draws conclusions and discusses future work.

\section{REINFORCEMENT LEARNING CONTROL}\label{sec:Methods}
    Here, we describe the algorithm used in this paper. First, we give a basic introduction to the Reinforcement Learning framework and formulate the AO control problem in this framework. After that, we will discuss the Deterministic Policy Gradient algorithm demonstrate how it can be applied to closed-loop AO control.
    
    \begin{table}[htbp]
    \caption{Overview and explanation of the various terms used in the RL framework.}
    \label{tab:terms}
    \center
    \begin{tabular}{l|l|l}
    \hline
    \textbf{Symbol}  &    \textbf{RL term}           & \textbf{AO term/Explanation}             \\ \hline 
    $s$ & State   & Input to the controller.         \\ 
    $o$  & Observation & Measurement of the residual wavefront.    \\ 
    $a$ &   Action   &  Incremental DM commands.                \\
    $r$ & Reward      & Measure of how good the current state is. This can be arbitrarily defined. \\
    $R$ & Return      & Discounted sum of future rewards, the optimization objective.    \\
    $\pi_\theta$ & Actor/policy     & The control law.                     \\ 
    $Q_\omega$ & Critic  &  Estimation of the return/the cost function. \\ 
    \end{tabular}
    \end{table}
    
    \subsection{AO control as a Markov Decision Process}
    Reinforcement Learning (RL) algorithms deal with Markov Decision Processes (MDPs), discrete-time stochastic control processes. In the RL framework, an agent operates within an environment, which is formally defined by a tuple $(\mathcal{S}, \mathcal{A}, \mathcal{R}, \mathcal{P})$ and time index $t$. Here, $s_t \in \mathcal{S}$ is the state of the environment, $a_t \in \mathcal{A}$ the action taken by the agent, $\mathcal{R}$ the reward function and $\mathcal{P}$ the state transition probabilities defined by:
    \begin{equation} \label{eq:markov}
       \mathcal{P}(s_t,a_t,s_{t+1}) = P(s_{t+1}| s_t, a_t),
    \end{equation}
    where $P(s_{t+1}| s_t, a_t)$ gives the probability of transitioning from state $s_t$ to state $s_{t+1}$ as a result of action $a_t$. Eq. \ref{eq:markov} is also known as the Markov property.
    
    In the case of AO control the agent is the DM controller. The environment consists of everything but the controller, so among others the evolution of the atmospheric turbulence and the DM dynamics. At each timestep the environment is in a state $s_t$. The representation of the state will be extensively discussed in section \ref{sec:state_representation}, but a first guess could be the most recent observation of the residual wavefront $o_t$. Based on this state, the controller takes an action $a_t$, following some control law $\pi$: $a_t= \pi(s_t)$. Since the controller operates in closed-loop, this action consists of the incremental voltages added to the DM, equivalent to an integral controller. The DM commands and the temporal evolution of the environment result in the transition to a new state $s_{t+1}$, according to the transition probabilities $\mathcal{P}$, and an accommodated reward $r_t = \mathcal{R}(s_t, a_t, s_{t+1})$. This reward is a measure of the performance of the controller and the reward function $\mathcal{R}$ can be arbitrarily chosen, as long as it can be calculated at every time step. The goal is to determine the control law $\pi$ that gives the highest cumulative future reward. To ensure this quantity is finite and to prioritize immediate rewards, future rewards are discounted at a rate $\gamma<1$, the discount factor. The discounted future return $R_t$ is then defined as:
    \begin{equation}
        R_t = \sum_{t=0} \gamma^t r_t = r_t+\gamma r_{t+1}+ \gamma^2 r_{t+2} + \dots\;.
    \end{equation}
    The optimization objective can then be formally defined as:
    \begin{equation}
        J(\pi) = \mathbb{E}_{s\sim \rho^\pi, a\sim \pi}[R].
    \end{equation}
    Here, $\mathbb{E}$ denotes the expectation value and $\rho^\pi$ the state visitation distribution following control law $\pi$. This state visitation distribution describes how often we end up in state $s$ if we follow control law $\pi$. In the rest of this work, when we denote an expectation value, it is always over $s \sim \rho^\pi$ and $a \sim \pi$ and we will leave this out for readability. There are a variety of algorithms to find an optimal control law in the above framework\cite{Arulkumaran2017_RL_review}. One of the most frequently used algorithms for continuous control problems is the (Deep) Deterministic Policy Gradient algorithm \cite{Silver_dpg, Lillicrap2015_ddpg}, which we will use here.
    
    \subsection{(Deep) Deterministic Policy Gradient}
    The (Deep) Deterministic Policy Gradient (DPG)\cite{Silver_dpg} algorithm uses two parameterized models:
    \begin{itemize}
        \item An \textbf{Actor} $\pi_\theta(s)$ that models the control law with parameters $\theta$. This model maps the state to the DM commands : $a_t =\pi_\theta(s_t)$.

        \item A \textbf{Critic} $Q_\omega(s,a)$ that models the expected return for a given state and DM command with parameters $\omega$: $Q_\omega(s_t,a_t) = \mathbb{E}[R_t|s_t,a_t]$. This is an estimation of the cost function that has to be optimized.
    \end{itemize}
    Both the actor and critic may be any kind of parameterized, differentiable model. These may for example be function approximators, such as Artificial Neural Networks (ANN) \cite{Goodfellow_deeplearning}. The DPG algorithm with ANNs as function approximators is referred to as the Deep Deterministic Policy Gradient (DDPG) algorithm \cite{Lillicrap2015_ddpg}. The algorithm finds the parameters $\theta$ of the controller that give the highest expected future return $R$. Using the actor and the critic we can calculate the gradients of the modelled expected return with respect to the control parameters using the chain rule:
    \begin{equation}
    \begin{split}
         \nabla_\theta J(\pi_\theta) =& \mathbb{E}[\nabla_\theta Q_\omega(s,a)]\\
         =& \mathbb{E}[\nabla_a Q_\omega(s,a) \nabla_\theta\pi_\theta(s)]
    \end{split}
    \end{equation}
    Using these gradients we can use a gradient-based optimizer to update the parameters $\theta$ of the control law in the direction in which the expected return increases. The quality of the controller thus inherently depends on the ability of the critic $Q_\omega(s,a)$ to successfully model the expected return. The critic can be taught using Temporal Difference (TD) learning, which uses the Bellman equation to bootstrap the expected future reward:
    \begin{equation}\label{eq:bellman}
        \begin{split}
    Q_\omega(s_t,a_t) = & \mathbb{E}[R_t| s_t, a_t]\\
    = &\mathbb{E}[r_t + \gamma r_{t+1} + \gamma^2 r_{t+2}+\dots] \\
              = &\mathbb{E}[r_t+\gamma (r_{t+1}+\gamma r_{t+2}+\dots)] \\
              = &\mathbb{E}[r_t+\gamma Q_\omega(s_{t+1},\pi(s_{t+1}))] \equiv \mathbb{E}[y_t],
        \end{split}
    \end{equation}
    where $y_t$ are referred to as the target values. Training the critic is then a supervised learning problem, where we have to find the parameters $\omega$ that minimize the mean-squared error between the output of the critic and the target values $y_t$. The critic loss $L$ is thus given by:
    \begin{equation}
        L(\omega) = \frac{1}{2}\mathbb{E}[(y_t - Q_\omega (s_t, a_t))^2],
    \end{equation}
   and its gradients with respect to the parameters $\omega$ by:
    \begin{equation}
        \nabla_\omega L (\omega) = \mathbb{E}[(y_t-Q_\omega (s_t,a_t))\nabla_\omega Q_\omega(s_t, a_t)].
    \end{equation}
    We can then use a gradient-based optimizer to update the parameters in the direction that minimizes the loss. However, since the target values depend on the critic itself, this may quickly lead to instabilities in the training process. It is therefore common to use target models $Q'$ and $\pi'$ to calculate the target values $y_t$ \cite{Lillicrap2015_ddpg}. The parameters of these target models slightly lag behind the true models. For every update of the critic, the target models are updated as:
    \begin{equation}
    \begin{split}
        \theta' = \tau \theta + (1-\tau)\theta' \\
        \omega' = \tau \omega + (1-\tau)\omega'.
    \end{split}
    \end{equation}
    Here, $\tau\ll 1$ is a hyper-parameter that determines how quickly the target models are updated.
    
    \subsection{Data collection}
    Updating the actor and critic requires evaluating expectation values over the state visitation distribution. This can naturally be done by collecting tuples ($s_t$, $a_t$, $r_t$ $s_{t+1}$) while following the control law $\pi_\theta$. This data collection process is often separated into episodes of finite length, after which the sequence of tuples is saved in a so-called replay buffer, which is our training dataset. For computational reasons, the expectation values are often estimated over a limited number of these tuples, a batch, instead of over the full dataset. To reduce the correlation in a batch the tuples are randomly sampled from the replay buffer. 
    
    The training data can also be generated using another controller than the current best controller. This can for example be done by randomly perturbing the DM commands during the data collection, which is called exploration in the RL framework. For example, we can add some zero mean, normally distributed exploration noise with standard deviation $\sigma$:
    \begin{equation}
        a_t = \pi_\theta(s_t) + \mathcal{N}(0, \sigma) 
    \end{equation}
    This results in the controller not always giving the same DM commands for a given state, such that we can observe the results of these different commands. A common trick is to use large initial exploration noise and slowly decay it. This allows the algorithm to first observe the results of large changes in DM commands and finally observe the result of small changes in these commands. We empirically choose the standard deviation of the exploration noise during episode $k$ as:
        \begin{equation}\label{eq:exploration}
        \sigma_k = \frac{\sigma_0}{1+\zeta k} \; \textrm{rad}.
    \end{equation}
    Here, $\sigma_0$ specifies the initial standard deviation of the exploration noise and $\zeta$ how quickly it decays. Something that we do not employ here, but is also possible, is to use a completely different controller for the data collection, such as an integrator. This will introduce a slight bias in the models, as the data is then not exactly distributed according to state visitation distribution $\rho^\pi$, but this can often be neglected.

    \subsection{State representation} \label{sec:state_representation}

    It is crucial to have a good representation of the state $s_t$. A first guess for the state might be the most recent wavefront observation, $o_t$. This may, for example, be retrieved from a dedicated wavefront sensor (WFS). However, the DDPG algorithm assumes that the environment obeys the Markov property (Eq. \ref{eq:markov}), which requires that the state-transition probabilities are fully described by the current state and the action taken. This is not trivial in closed-loop AO control. First of all, we do not have access to the full state of the environment but only observe a noisy representation with the WFS measurements. Furthermore, the most recent wavefront measurement does not contain all the required information; for example, it lacks information about possible vibrations and wind flow. When the AO loop is closed, we also have to distinguish changes in the turbulence profile from changes due to previous commands, as we only observe the closed-loop residuals. These issues may be solved using state augmentation, where sequences of previous wavefront measurements and applied DM commands are used in the state. However, it may require a large number of previous measurements to have a full representation of the state. For example, Guyon \& Males\cite{Guyon2017_eof} used a vector of the previous 800 measurements to correct tip-tilt vibrations. A disadvantage of this is that it increases the computational demand. Furthermore, all previous measurements are in principle weighted equally, which may lead to additional noise if there are spurious correlations in the training data.
%CUK: You could add something about the number of required measurements to detect vibrations of a certain frequency. In the classical case, you would need to cover at least a whole period. I am not sure what you need in your case, but an autoregressive model needs very little to time coverage to detect long-period oscillations.
     
     To circumvent the problem of partial observability, it was shown that Recurrent Neural Networks (RNNs) are able to solve control problems even for partially observable MDPs (POMDPs), where the Markov property is not obeyed \cite{Wierstra2007, Heess2015_rdpg}. RNNs are a type of neural network often used in processing sequential data. Instead of only mapping an input to an output, RNNs also update a memory state $m_t$ based on the previous memory state and the input. This allows it to summarize previous information in this memory state and then use it to calculate the output. This way, RNNs can learn an internal representation of the true state of the system based on sequences of noisy measurements. An additional advantage is that calculating the control commands only requires the most recent state to be propagated through the model. This is computationally less demanding, which may be important for high-order AO systems operating at kHz frequencies. Furthermore, RNNs also effectively use the temporal structure of the data by incorporating basic priors, such as that the most recent measurement is the most relevant instead of using one large vector of components with equal weights. In our case, we will use a Long Short-Term Memory (LSTM) cell \cite{Hochreiter1997_lstm} as recurrent architecture. Training is now done on sequences of state-action-reward tuples using Truncated Backpropagation Trough Time (TBTT) over a length $l$. This version of the algorithm is known as the Recurrent Deterministic Policy Gradient (RDPG) algorithm \cite{Heess2015_rdpg}.
     
     In this case, the state for closed-loop AO can consist of only the most recent observation of the wavefront $o_t$ and the previous DM commands $a_{t-1}$:
     
     \begin{equation}
         s_t = (o_t, a_{t-1})\;.
     \end{equation}

    \subsection{Algorithm Overview}
    In summary, the algorithm can be described by three main steps that are iteratively performed:
    \begin{enumerate}
        \item Collect training sequences ($o_1,a_1,r_1,\dots o_{t}, a_t, r_t$) by running closed-loop and storing the data in the replay buffer.
        \item Train the critic using TD learning on the observed data.
        \item Improve the control law using the gradients obtained from the critic.
    \end{enumerate}
    The pseudo-code describing the full algorithm can be found in Algorithm \ref{algo:rdpg}.

    \RestyleAlgo{ruled}
    \begin{algorithm}[htbp]
     \caption{Recurrent Deterministic Policy Gradient for closed-loop AO (based on Heess et al. \cite{Heess2015_rdpg})}
      \label{algo:rdpg}
     Initialize critic parameters $\omega$ and actor parameters $\theta$ randomly\\
     Initialize target network parameters $\omega'=\omega$, $\theta'=\theta$ \\
     Initialize replay buffer \\
     Initialize $\sigma_0$ \\
     \For{k=1,number of episodes}{
       Reset controller memory states $m_t=0$ and deformable mirror \\
       \For{t=1,T}{
         Receive wavefront measurement $o_t$, reward $r_t$, construct state $s_t = (o_t, a_{t-1})$ and select incremental DM commands $a_t = \pi(s_t) + \epsilon$, with $\epsilon \sim \mathcal{N}(0,\sigma_k)$
      }
      Save trajectories $(o_1,a_1,r_1,\dots,o_T,a_T,r_T)$ in replay buffer \\
      Decay exploration noise $\sigma_k = \frac{\sigma_0}{1+\zeta k}$\\
      \For{j=1,number of training steps}{
      Randomly sample N batches of length $l$ from the replay buffer \\
      For each batch $i$ construct state histories $h_t^i = ( (o_1^i, a_0^i) \dots, (o_t^i, a_{t-1}^i))$ \\
      Calculate the critic targets using the Bellman equation: $$y_t^i = r_t^i + \gamma Q'_{\omega'}(h_{t+1}^i,\pi'_{\theta'}(h_{t+1}^i)) $$\\
      Calculate the sampled critic loss gradient with TBTT:
      $$ \nabla_\omega L(\omega) \approx \frac{1}{Nl} \sum_{i=1}^N \sum_{t=1}^l (y_t^i - Q_\omega(h_t^i,a_t^i))\nabla_\omega Q_\omega(s_t^i,a_t^i) $$\\
      Calculate the sampled Policy Gradient with TBTT:

      $$\nabla_\theta J(\theta) \approx \frac{1}{Nl} \sum_i^N \sum_t^l \nabla_{a} Q_\omega(h_t^i,a_t^i) \nabla_\theta \pi_\theta(h_t^i) $$\\
      Use a gradient-based optimizer (e.g. Adam) to update the actor parameters $\theta$ and critic parameters $\omega$.\\
      Update the target network parameters:
      \begin{align*}
        \theta' &= \tau \theta + (1-\tau)\theta' \\
        \omega' &= \tau \omega + (1-\tau) \omega' \\
      \end{align*}

      }
      }
    \end{algorithm}
    
\section{Simulation results} \label{sec:tip_tilt_simulations}
    \subsection{Simulation setup}
    As a proof of concept we apply the algorithm to tip-tilt control in simulations. We simulate an idealized optical system using the hcipy \cite{por2018hcipy} package for python. Our system has an unobscured, circular aperture of 1 meter in diameter and operates at a wavelength of 1~$\mu$m. The observation $o_t$ is the center of gravity ($x_t, y_t$) of the focal plane image in units of $\lambda/D$. The measured center of gravity along with the previous DM commands $a_{t-1}$ are fed into a controller that controls tip and tilt in closed loop. To represent the servo-lag of estimating the wavefront tip and tilt, calculating and applying the control commands, we delay the DM commands by a discrete number of frames:
    \begin{equation}
        \textrm{DM}_t = \textrm{DM}_{t-1} + a_{t-\tau},
    \end{equation}
    where $\tau$ is the servo lag. This is equivalent to the controller seeing the state from $\tau$ iterations ago, which implies a delay in obtaining the reward. Throughout this section we assume an AO system operating at 1 kHz with a servo lag of 3 frames and do not consider any detector noise. Furthermore, we only consider tip-tilt errors and ignore all higher order modes in the simulations.
    
    \subsubsection{Algorithm setup} \label{sec:hyperparams}
    To minimize the residual tip-tilt jitter we choose the squared PSF center deviation as the reward function:
    \begin{equation}
        r_t = -\frac{x_t^2+y_t^2}{b^2}.
    \end{equation}
    Here, $b$ is a scaling factor for which we use a value of $4 \lambda/D$ in the simulations. We use the same Neural Network architecture for both the actor and the critic. The input consists of the wavefront measurement $o_t= (x_t, y_t)$ and the previously applied tip-tilt commands $a_t = (a_{x,t-1}, a_{y,t-1})$. We use a Long Short-Term Memory \cite{Hochreiter1997_lstm} (LSTM) cell with 64 neurons as the recurrent layer in both the actor and the critic. For the critic the DM commands $a_t$ are appended to the output of the LSTM. After that, we have a fully connected layer with 64 neurons for both networks. Finally, the output of the actor consists of two neurons, the additional tip and tilt commands, while the output of the critic is a single neuron that estimates the expected future return $Q(s_t, a_t)$. The LSTM uses a tanh activation function for the input and output gates and a hard sigmoid for the recurrent gates while the fully connected layer uses the ReLU activation function. The output of the actor again uses a tanh activation function to constrain the incremental DM commands between -1 rad and +1 rad.
    The output of the critic has a linear activation function. We use the gradient-based Adam optimizer algorithm \cite{Kingma2015_adam} with default parameters except for the learning rate. After every 500 iterations (an episode), we reset the input turbulence and DM shape. To reduce the correlation of the data in a batch early on, we only start training after a certain number of episodes, the warmup. We use a warmup of 5 episodes, which is equivalent to 2500 iterations, and the exploration law given in Eq. \ref{eq:exploration}. The hyperparameters of the RDPG algorithm are given in Table \ref{tab:hyperparams}.
    
    \begin{table}[htbp]
  \center
\caption{Hyperparameters}
\label{tab:hyperparams}
\begin{tabular}{c|c}
\hline
Parameter                         & Value          \\ \hline
Actor learning rate               & $10^{-5}$      \\
Critic learning rate              & $10^{-3}$      \\
Target network soft update $\tau$ & $10^{-3}$      \\
Discount factor $\gamma$          & $0.99$         \\
Batch size                        & 64             \\
TBTT length $l$                   & 50             \\
Initial exploration $\sigma_0$      & 0.3 rad       \\
Exploration decay $\zeta$         & 0.005         \\
Episode length                    & 500 iterations \\
Number of training steps per episode& 500 \\
\end{tabular}
\end{table}
    
    \subsection{Vibration Suppression}
\begin{figure}[htbp]
            \centering
            \includegraphics[width=0.46\linewidth]{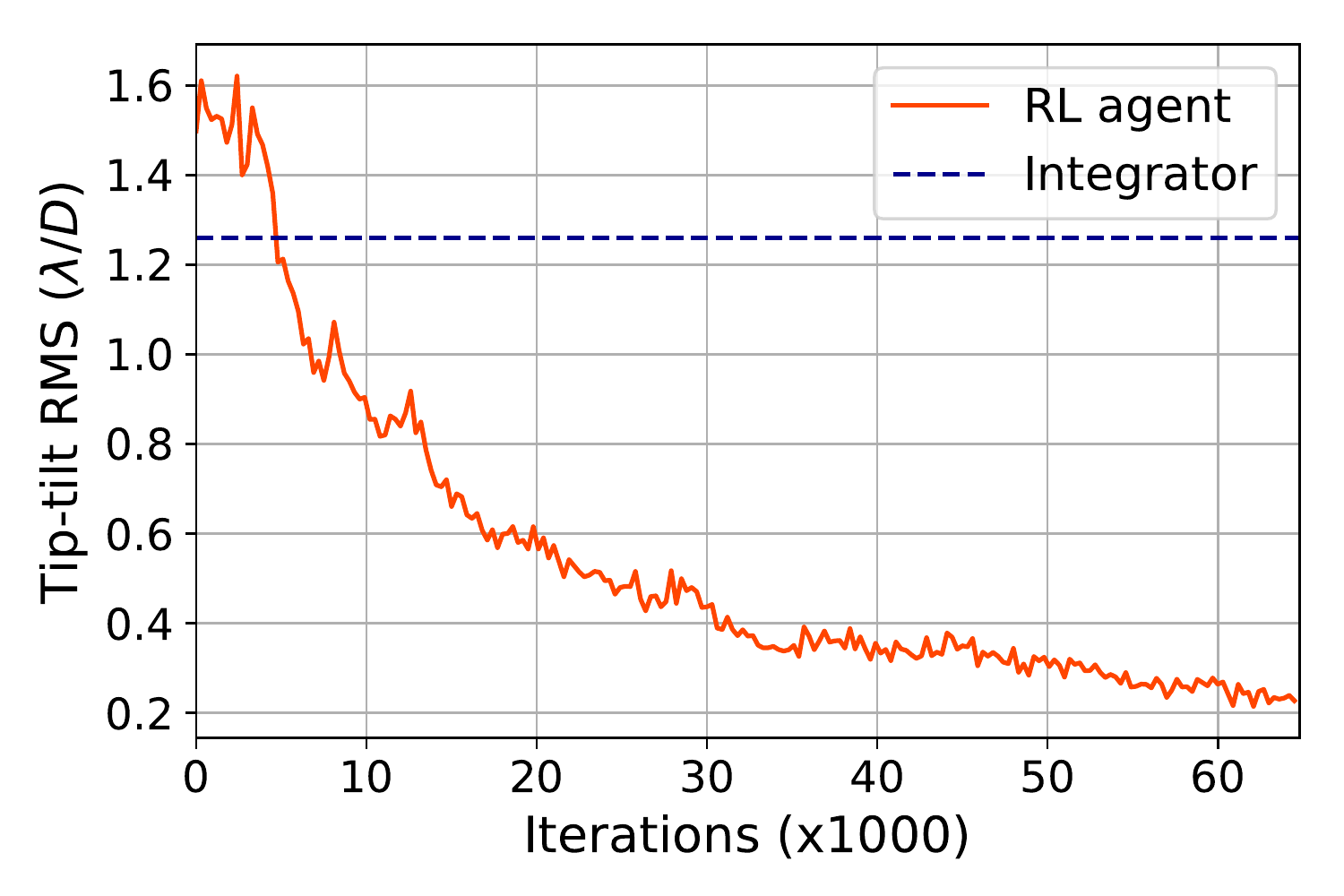}
            \includegraphics[width=0.46\linewidth]{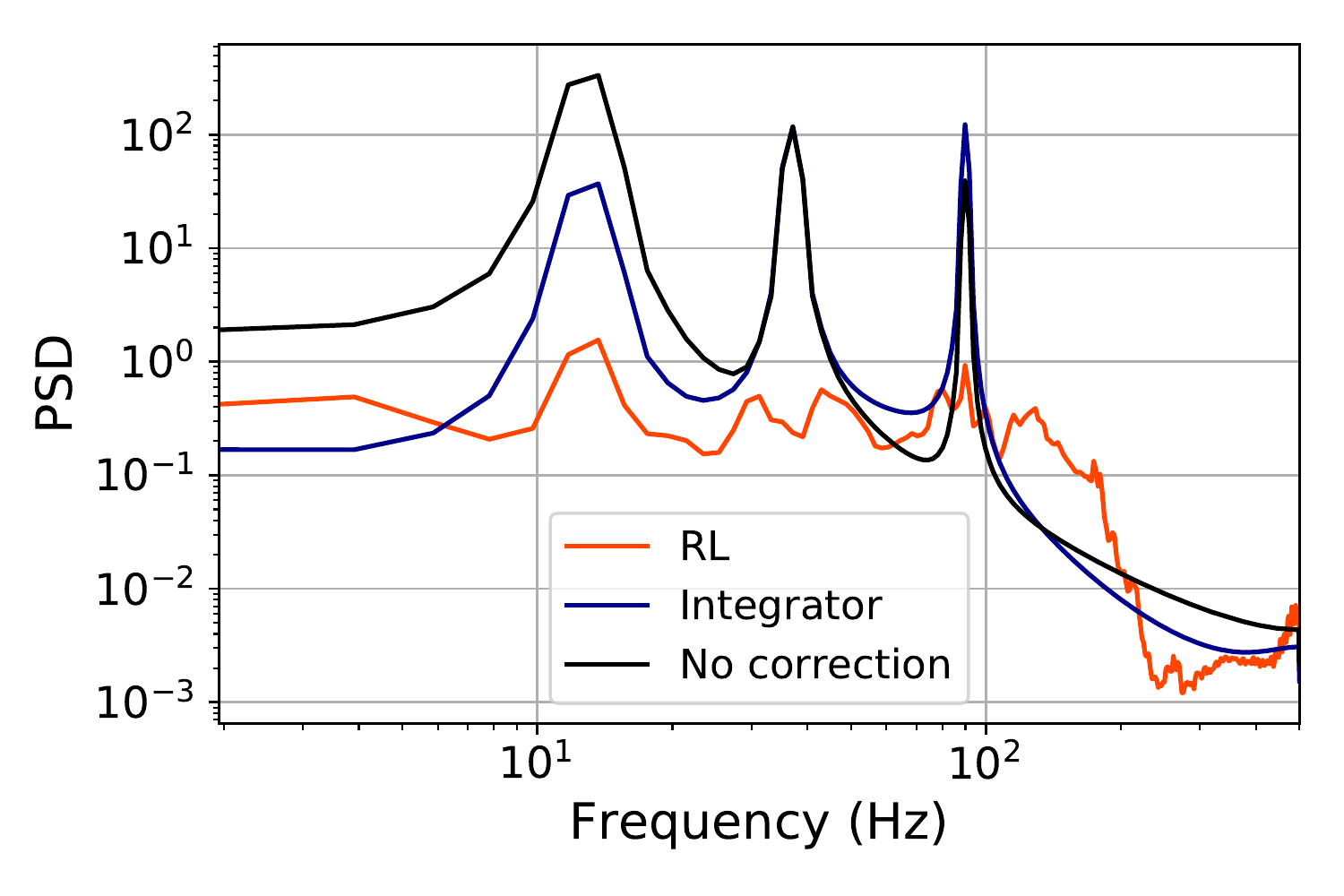}
            \includegraphics[width=0.68\linewidth]{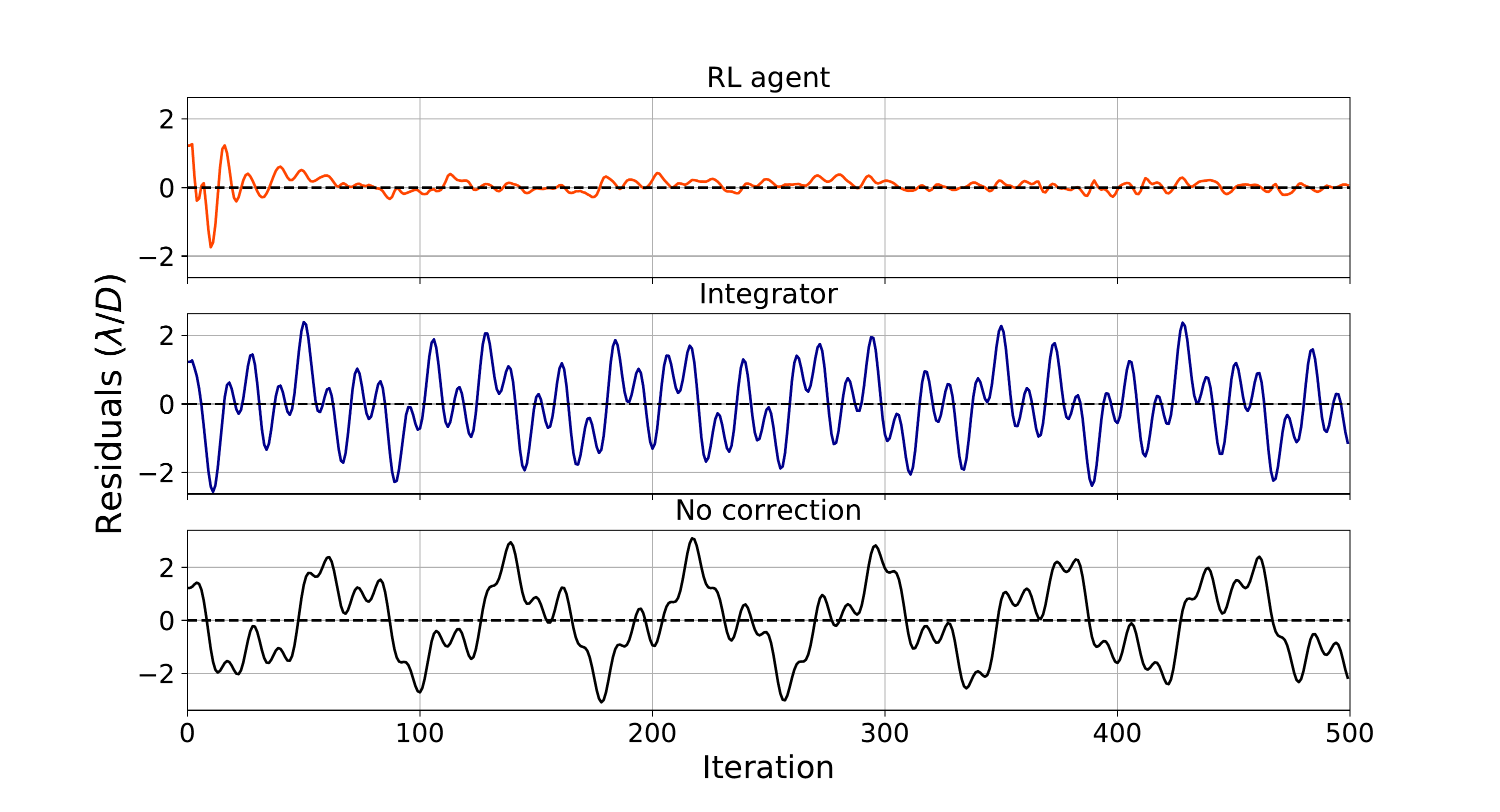}
            \caption{\textbf{Top left:} Training curve of the Reinforcement Learning controller under input turbulences consisting of three pure vibrations along both directions. Also shown is the average performance of an integrator with optimal gain. \textbf{Top right:} Temporal PSD of the residuals in the $x$-direction for a simulation of 10 seconds. Note that the lines of the input spectrum and integrator overlap at the peak at 37 Hz. \textbf{Bottom:} Residuals for the different controllers in the $x$-direction for a 500 ms simulation.}
            \label{fig:vibrations}
        \end{figure}

    Here we consider input turbulence consisting of three pure vibrations along both the $x$ and $y$ directions. Along the $x$ direction we have vibrations with frequencies at 13 Hz, 37 Hz and 91 Hz and along the $y$ direction at 11 Hz, 43 Hz and 87 Hz. The gain of the integrator is optimized by running in closed loop for 1 second and choosing the gain that gives the lowest residual root mean square (RMS) center deviation:
    \begin{equation}
        \textrm{RMS} = \sqrt{<x_t^2+y_t^2>}.
    \end{equation}
    Figure \ref{fig:vibrations} shows the evolution of the RMS during the training of the RL controller. It shows that after a few thousand iterations the RL controller outperforms the integrator and eventually reaches an average RMS that is a factor ~6 lower than for the integrator. The figure also shows the temporal PSD of the residuals along the $x$-direction for the fully trained controller without exploration noise. This PSD is estimated using Welch's method\cite{Welch1975} for a simulation of 10 seconds. The integrator reduces the power of the frequency at 13 Hz, but the other vibrations fall outside its correction bandwidth, even adding power to the 91 Hz vibration. The RL controller almost completely removes the power from all three vibrations. The most power is left at a frequency of 13 Hz. This may be explained by the fact that the period of this vibration is longer than the optimization length of 50 ms. The residuals in the $x$-direction for a 500-ms simulation for the RL controller, integrator and input spectrum are shown in Fig. \ref{fig:vibrations}.
        {}
    
    \subsection{Power-law Disturbances}
\begin{figure}[htbp]
            \centering
            \includegraphics[width=0.49\linewidth]{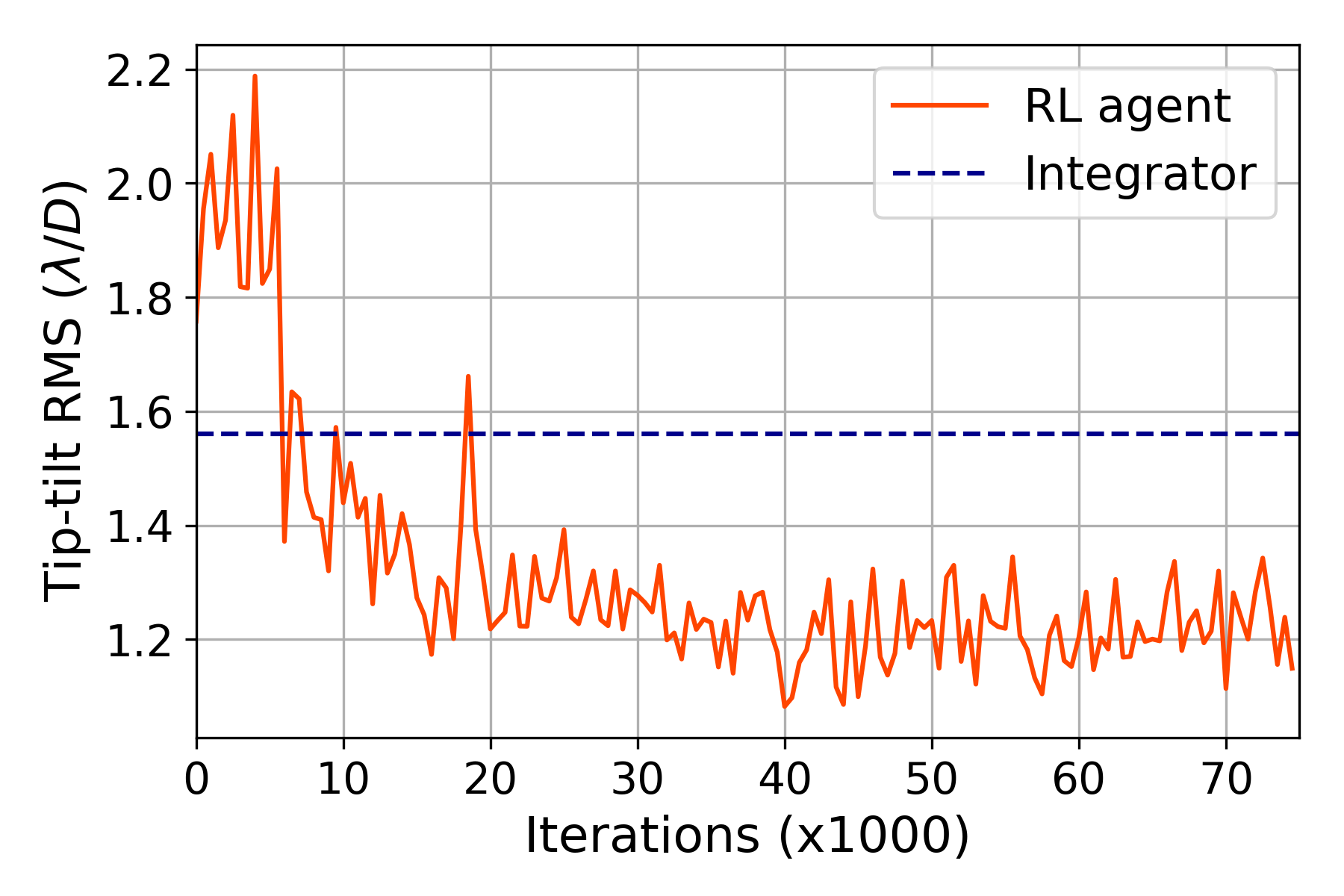}
            \includegraphics[width=0.5\linewidth]{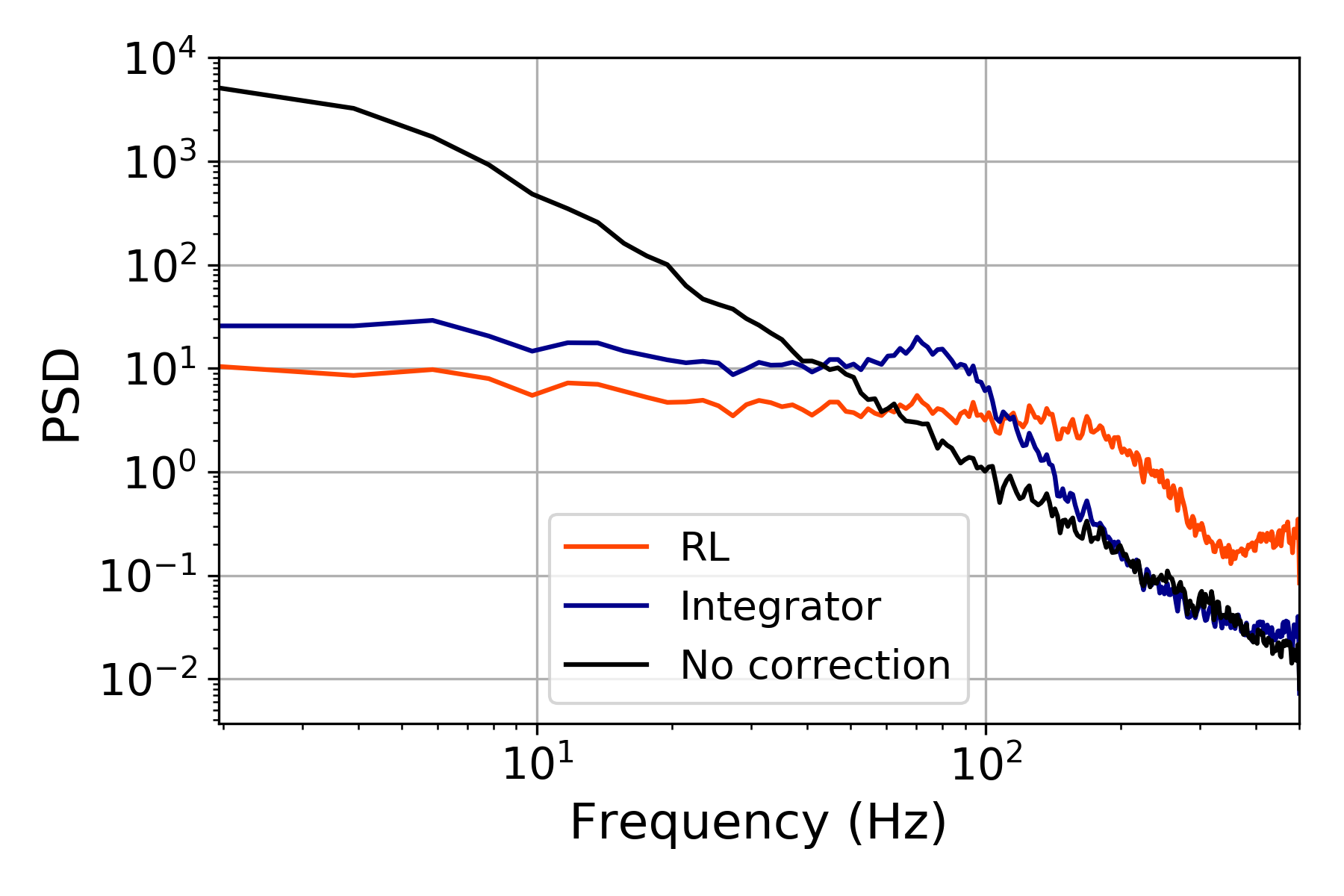}
            \caption{\textbf{Left:} Training curve of the Reinforcement Learning controller under power-law input turbulence and the average performance of an integrator with optimal gain. \textbf{Right:} Temporal PSD of the residuals in the x-direction for a simulation of 10 seconds.}
            \label{fig:power_law}
        \end{figure}

    Next, we consider input disturbances following a temporal power law with a power-law index of -8/3 along both directions. For every episode, we generate random power-law time series for 500 ms, neglecting frequencies lower than 2 Hz. The training curve is shown in Fig. \ref{fig:power_law}. We again observe better RMS performance for the RL controller as compared to an integrator with optimized gain. We test the trained controller by running in closed loop for 10 s and calculating the temporal PSD of the residuals, which are also shown in Fig. \ref{fig:power_law}. It shows that the RL controller has a larger rejection power at lower frequencies. The residual PSD of the integrator exhibits a bump around ~70 Hz, which is the result of the integrator overshooting. A lower gain would reduce this overshooting but would also decrease its rejection power at lower frequencies. The RL controller can account for already applied commands that are not yet visible in the residual measurements due to the servo lag, allowing it to use a higher gain without overshooting.
    
        {}

\section{LAB RESULTS}\label{sec:tip_tilt_lab}
 \subsection{Experimental Setup}
\begin{figure}[htbp]
        \centering
        \includegraphics[width=0.9\linewidth]{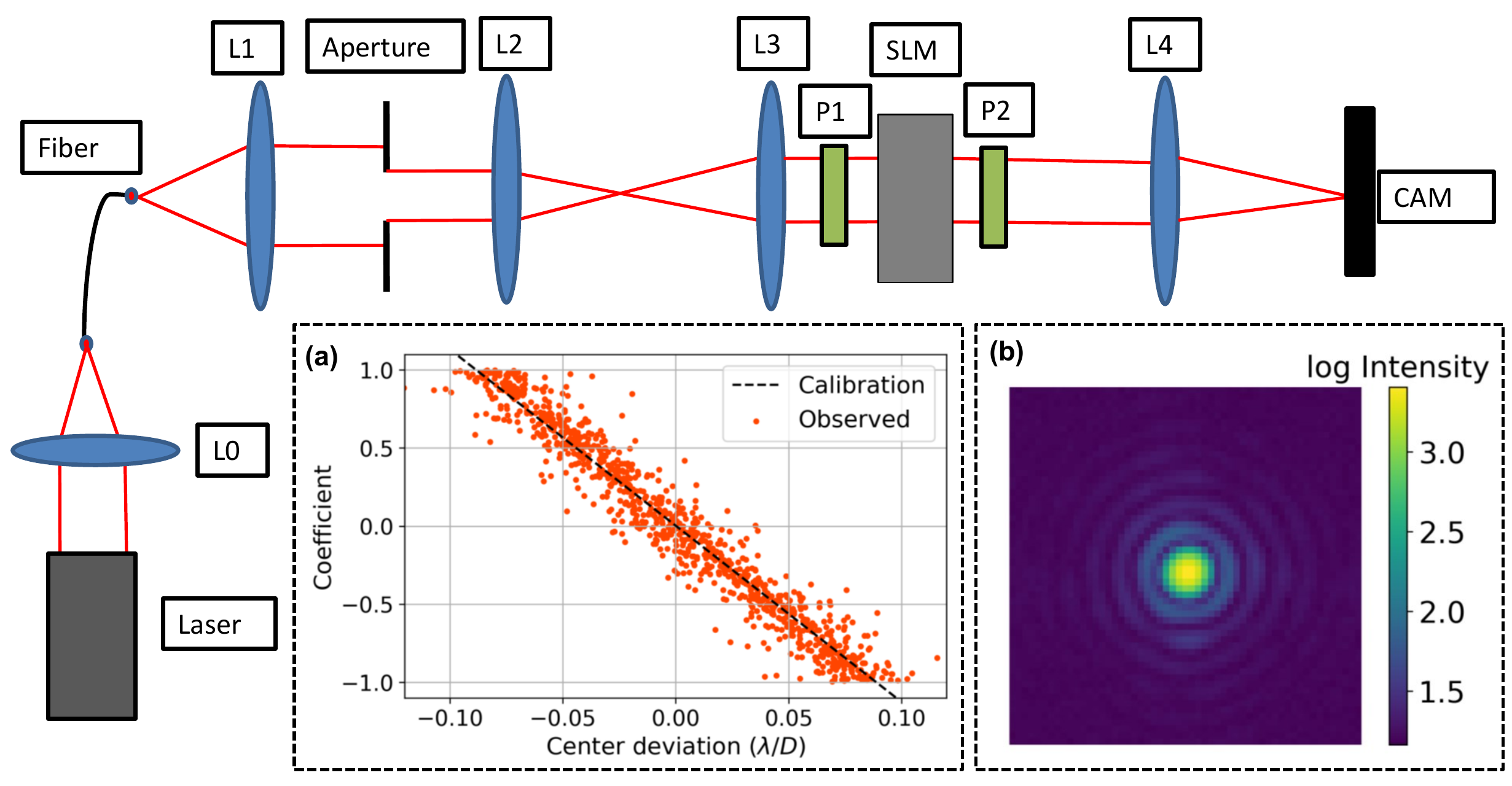}
        \caption{Lab setup to test the RL control algorithm with lenses L0-L4, Polarizers P1 and P2, a Spatial Light Modulator (SLM) and a camera (CAM). Insets: \textbf{(a)} Measured tilt in the focal plane as a function of the applied phase gradient on the SLM. \textbf{(b)} Measured Point Spread Function for the setup.}
        \label{fig:lab_setup}
    \end{figure}

The results in the previous section are for idealized simulations. To validate our results for tip-tilt control we now test the algorithm in the lab. The lab setup is sketched in Figure \ref{fig:lab_setup}. The first lens focuses a 633-nm He-Ne laser onto a single-mode fiber. The fiber output is then collimated, and a diaphragm defines the input aperture. A 4f system reimages the aperture onto a Holoeye LC2002 Spatial Light Modulator (SLM) with 800x600 pixels. The pupil covers approximately 200x200 pixels on the SLM. Right before the SLM we define the polarization with a polarizer. After a second polarizer we focus the beam onto the camera. The resulting Point Spread Function is also shown in Figure \ref{fig:lab_setup}.

We use the SLM as both the turbulence generator and the corrector. The polarizers are rotated such that the SLM mainly manipulates the phase profile with minimal amplitude modulation. The voltages of the SLM are controlled with 8-bit resolution. As the response for both small and large voltages becomes highly nonlinear, only values between 25 and 225 are used. Tip and tilt aberrations are introduced by applying a gradient on the SLM at the location of the pupil. Figure \ref{fig:lab_setup} shows the calibration of the SLM for the resulting shift of the PSF in the focal plane image. The coefficient refers to the amplitude of the gradient profile, where 1 corresponds to the maximum possible gradient in the pupil. We see that the dynamic range is a shift of approximately 0.1 $\lambda/D$. The refresh rate of the SLM is 30 Hz according to the specifications of the manufacturer. We use the SLM at a frequency of approximately 18 Hz. We have attempted to measure the delay of the control loop but this appeared to not be constant. This might be the result of varying processing speeds in the pipeline. On average, there is a delay of 0.1 to 0.15 seconds, which is equivalent to 2-3 frames at 18 Hz. For the RL algorithm we use the same architectures, hyperparameters and reward function as in the simulations (see Section \ref{sec:hyperparams}), except for the reward scaling $b$, for which we now use a value of 0.05 $\lambda/D$.

\subsection{Vibration Suppression}
\begin{figure}[htbp]
        \centering
        \includegraphics[width=0.5\linewidth]{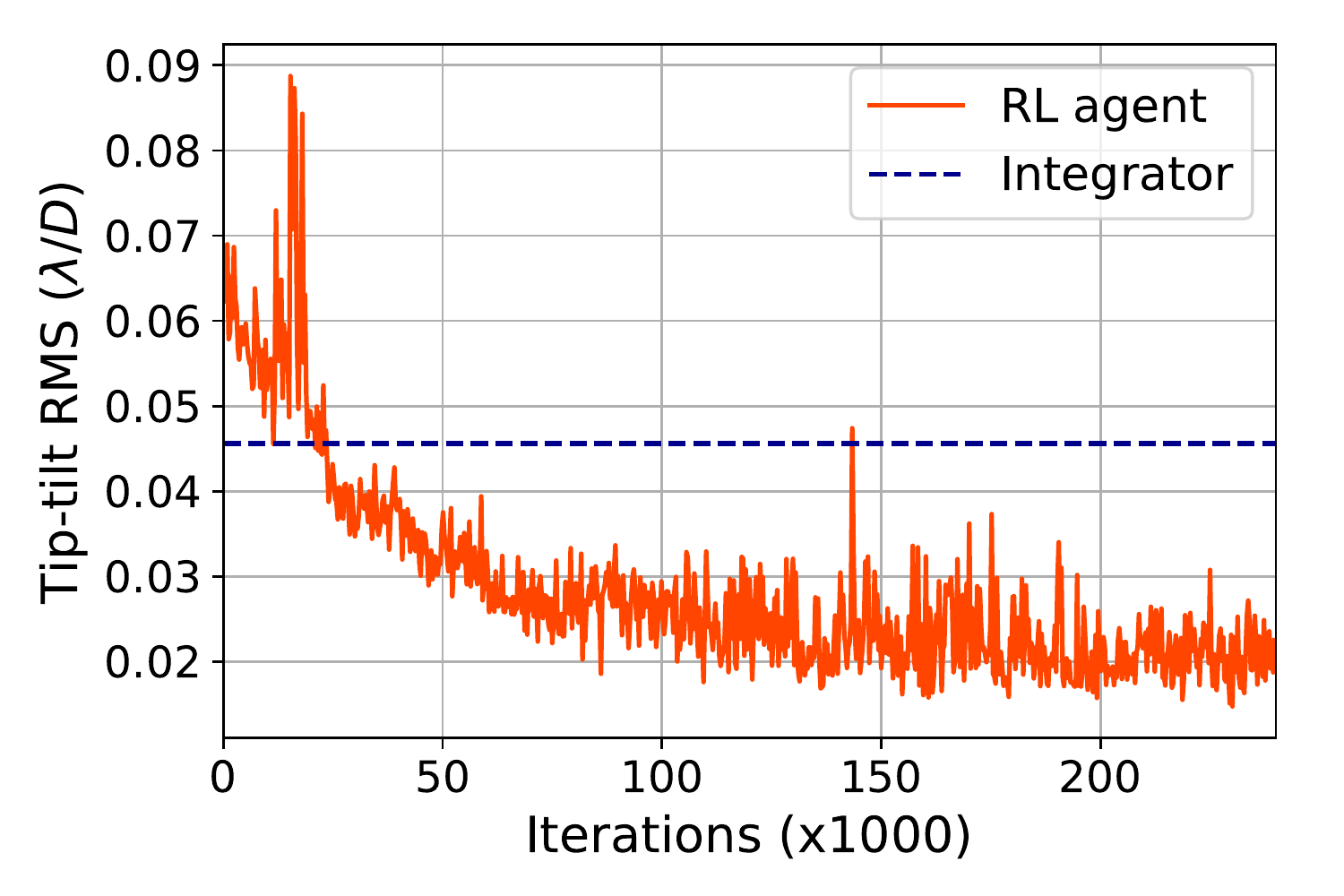}
        \includegraphics[width=0.49\linewidth]{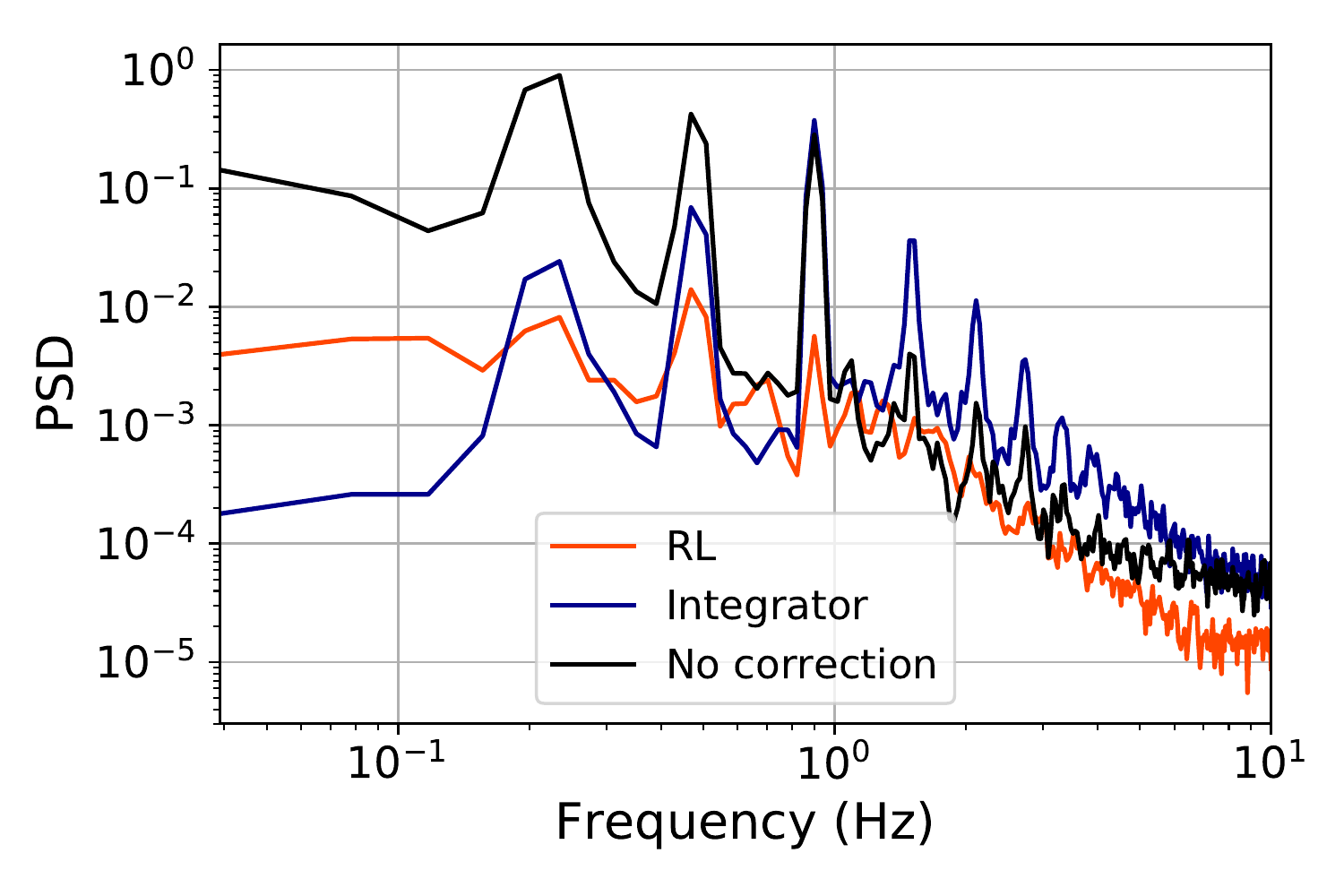}
        \includegraphics[width=0.7\linewidth]{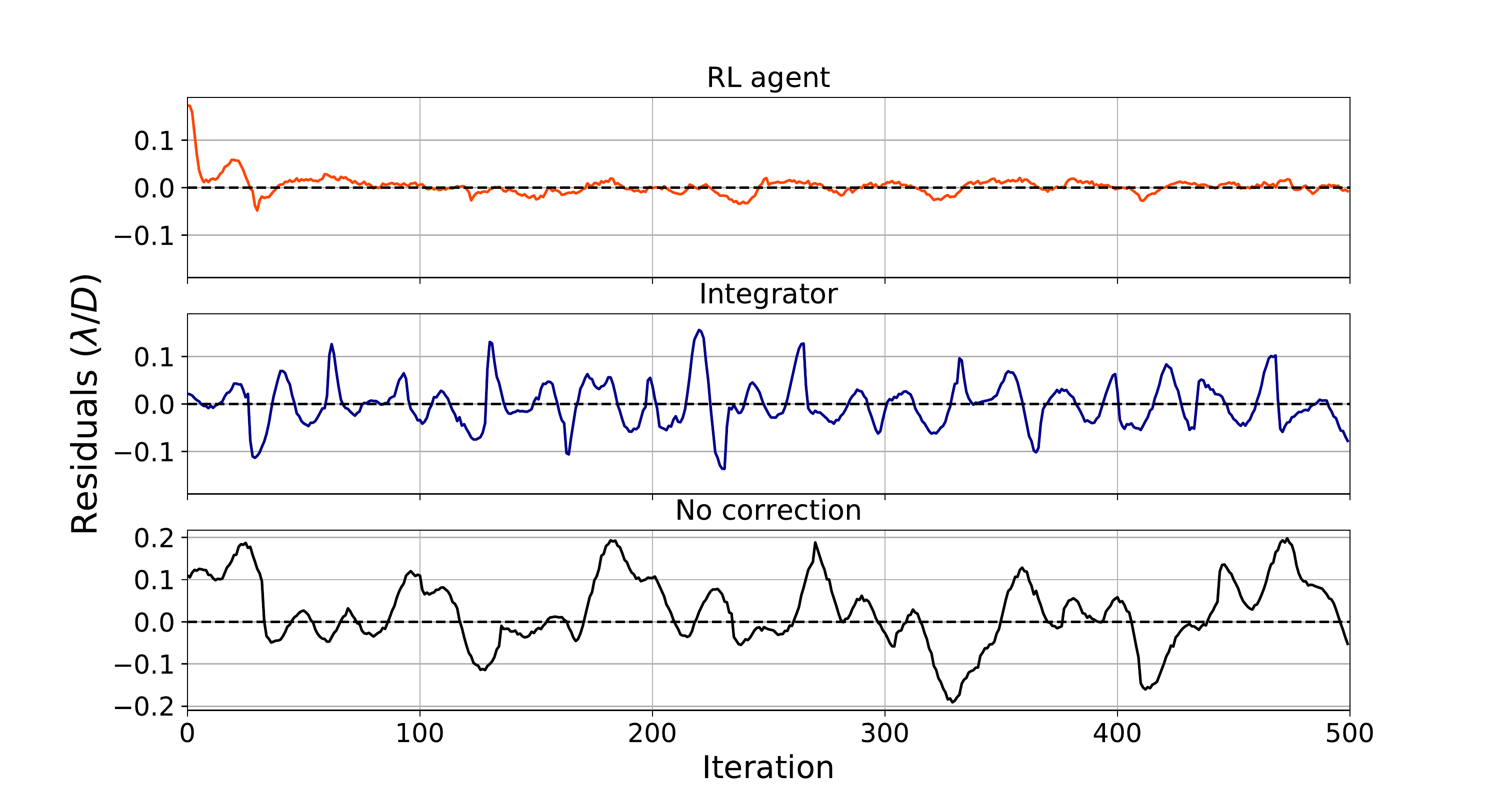}
        \caption{\textbf{Top left:} Training curve of the Reinforcement Learning controller for an input spectrum consisting of 3 vibrations along both directions in the lab. Also shown is the average performance of an integrator with optimal gain. \textbf{Top right:} Temporal PSD of the residuals in the x-direction for a measurement of 2500 iterations. \textbf{Bottom:} Residuals for the different controllers in the $x$-direction for a 500-iteration measurement. }
        \label{fig:lab_vibrations}
    \end{figure}

We again test the ability of the algorithm to filter out a combination of 3 vibrations along both directions. Figure \ref{fig:lab_vibrations} shows the training curve and the residual temporal PSD for the trained agent and an integrator with optimized gain. It also shows the residuals for a 500-iteration measurement. These results confirm our findings in the simulations as we again see a much reduced power in the vibrations. There is a little more power left in the vibrations as compared to the simulations, and the training time is longer. This is likely the result of the additional noise in the lab, variable control-loop delay and dynamics of the SLM. 
    {}
    
\subsection{Identification and mitigation of a varying vibration}
\begin{figure}[htbp]\label{fig:varying_vib}
    \centering
    \includegraphics[width=0.5\linewidth]{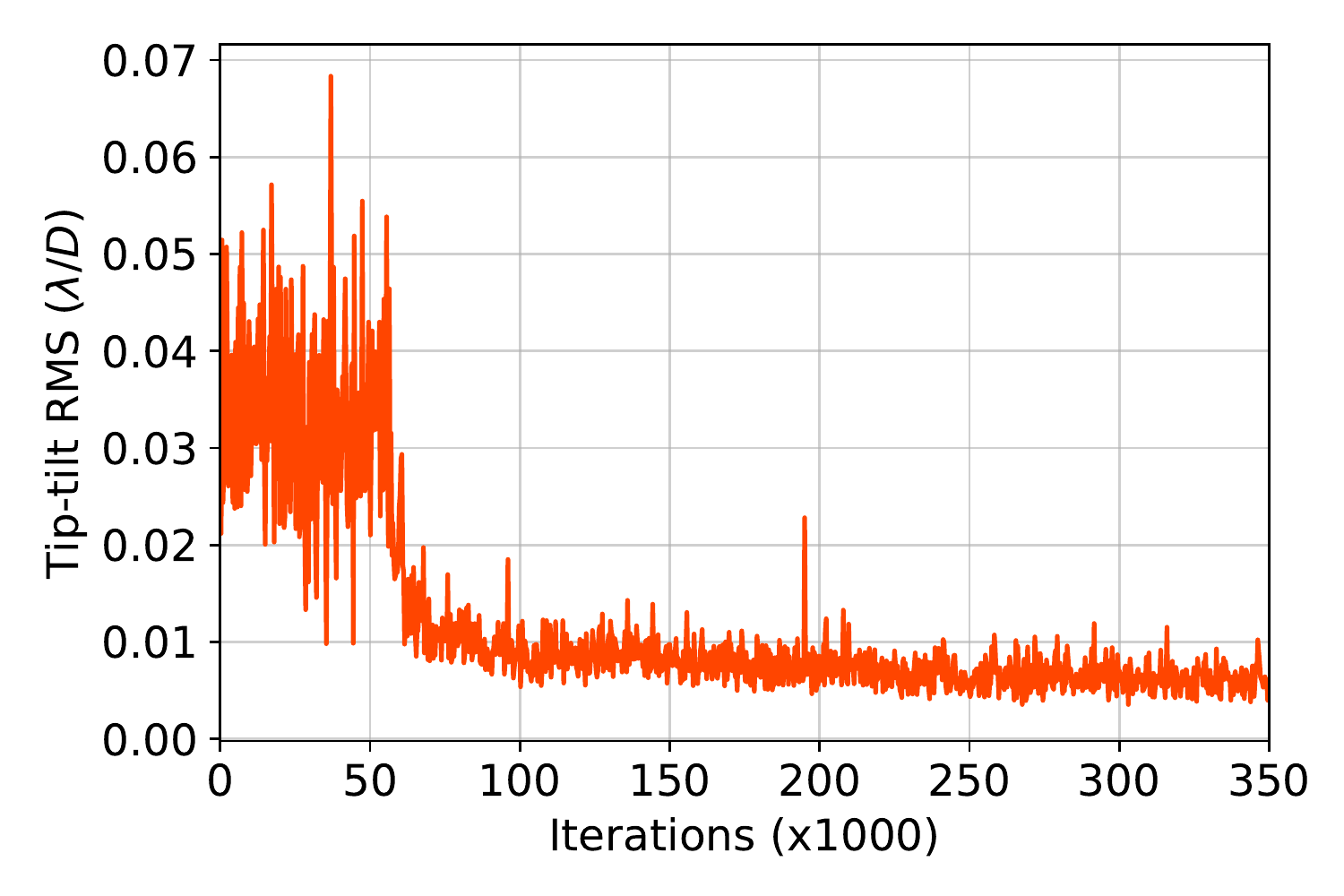}
    \includegraphics[width=0.49\linewidth]{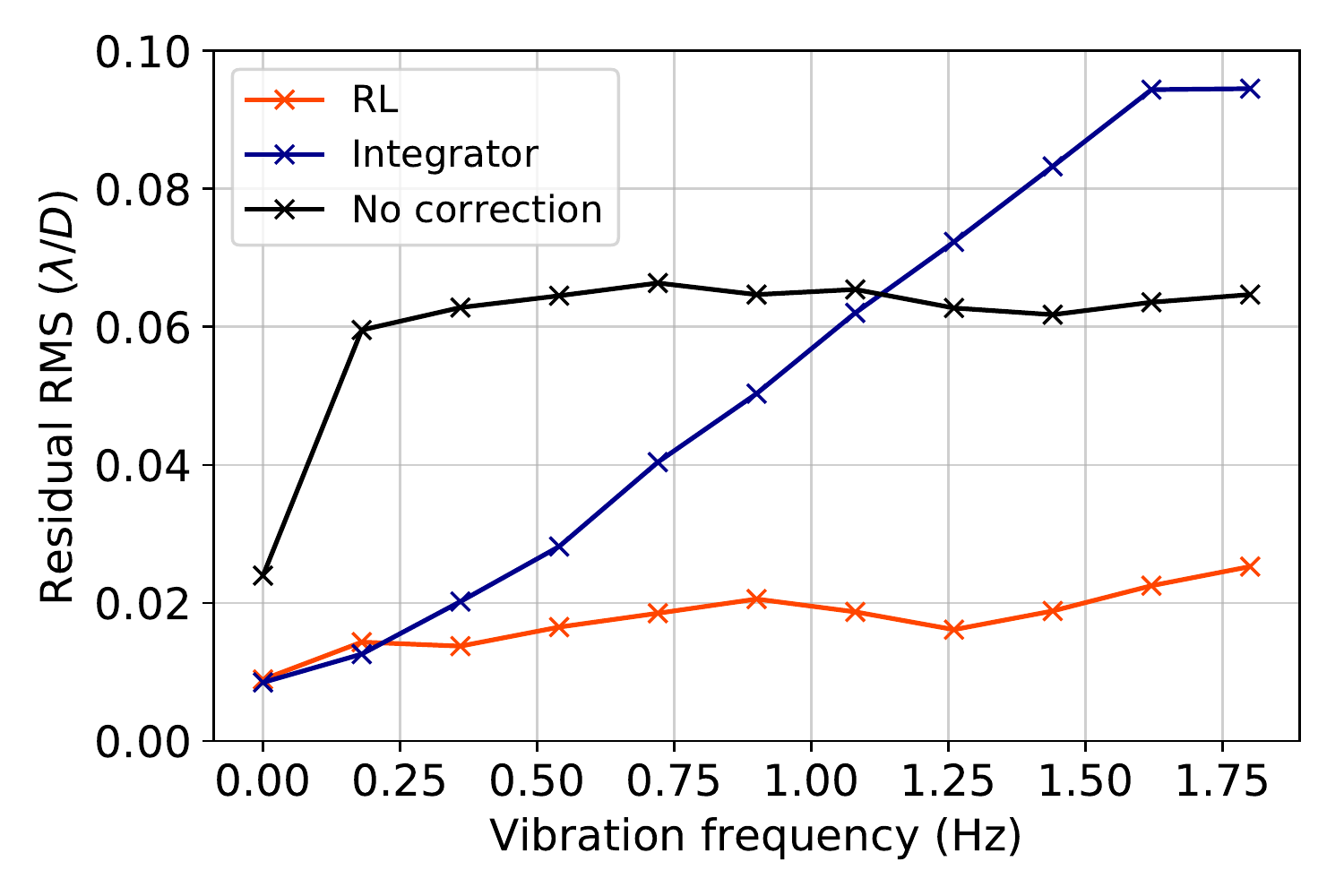}
    \caption{\textbf{Left:} Training curve of the Reinforcement Learning controller for a single vibration with random amplitude between 0 and 1 and random frequency between 0 and 1.8 Hz. \textbf{Right:} Residual RMS as a function of vibration frequency for a 2000-iteration measurement.}
    \label{fig:my_label}
\end{figure}

So far, we have only considered stationary input spectra. Once the input spectrum changes (e.g. the frequency of the vibrations changes), we would have to retrain the controller. However, Recurrent Neural Networks should be able to identify the relevant parameters online, without retraining. To test this, we train the controller with a single but randomly varying input vibration along the $x$-axis. After every 500 iterations we reset the vibration parameters: the amplitude is randomly sampled between 0 and 1 and the frequency between 0 and 1.8 Hz. The training curve is shown in Fig. \ref{fig:varying_vib}. Once it has converged there is very little variance in the RMS, meaning that the performance is mostly independent of the amplitude and frequency of the input vibration. We compare the performance of the trained controller to that of an integrator with a fixed gain of 0.6. We apply a vibration with a fixed amplitude of 1 and a specific frequency and run closed loop for 2000 iterations. The residual RMS as a function of the frequency of the vibration is shown in Fig. \ref{fig:varying_vib}. We see that the RL controller is able to reduce the RMS for all frequencies. On the other hand, the integrator amplifies the vibrations starting at a frequency of 1.25 Hz. This is the result of the vibration falling outside the correction bandwidth for this gain. However, the RNN is able to adaptively change its gain based on the observed sequence of measurements. This demonstrates that the RNN can learn to identify the relevant parameters of the turbulence, in this case the amplitude and frequency of the vibration, without updating the control law.
%CUK: not sure what 0Hz means; it should be a constant, and as such the RMS should be zero --> This is without applying a vibration, so the lab noise spectrum.
\newpage
    
\subsection{Power-law Disturbances}
\begin{figure}[htbp]
        \centering
        \includegraphics[width=0.5\linewidth]{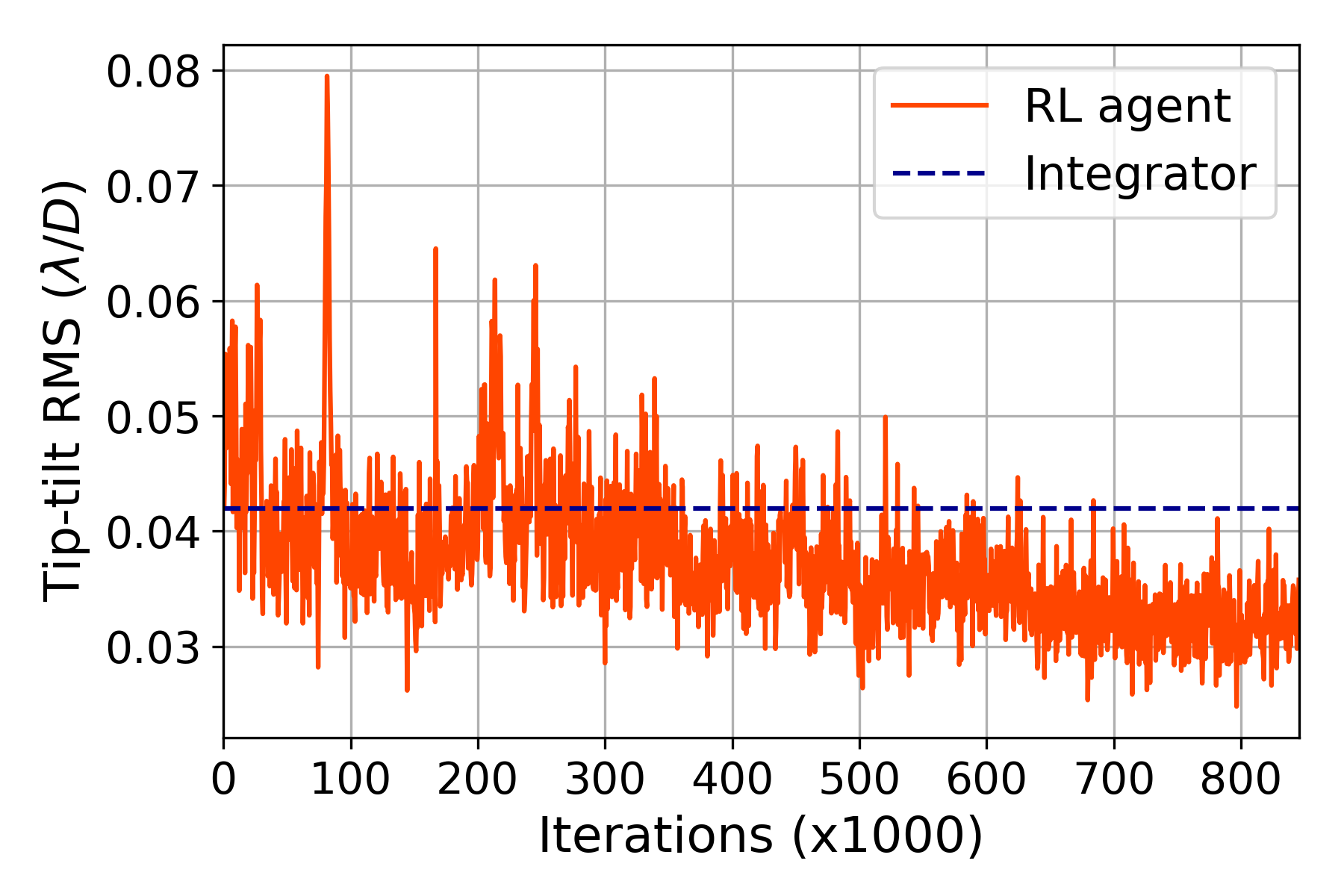}
        \includegraphics[width=0.49\linewidth]{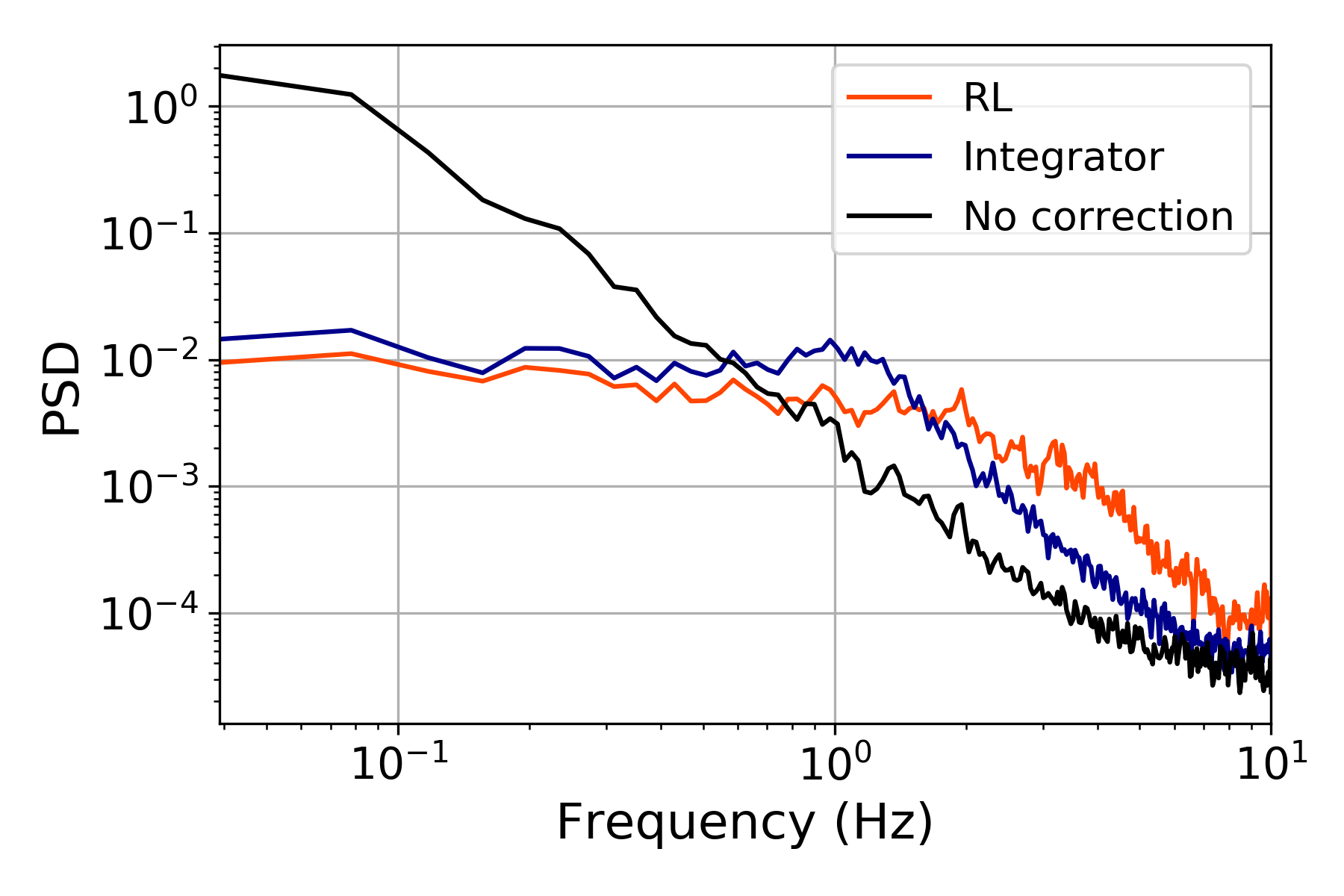}
        \caption{\textbf{Left:} Training curve of the Reinforcement Learning controller under power-law input turbulence in the lab. Also shown is the average performance of an integrator with optimal gain. \textbf{Right:} Temporal PSD of the residuals in the x-direction for a measurement of 2500 iterations.}
        \label{fig:lab_power}
    \end{figure}

We also test the performance for a power-law input spectrum. The training curve and residual temporal PSD are shown in Fig. \ref{fig:lab_power}. It takes many iterations before it converges. This is likely because of the noisy response of the SLM. Comparing the residual PSD to that of an integrator with optimized gain, we see results that are similar to the simulations, although the improvement is smaller. We again observe an increased rejection power at low frequencies without the bump due to overshooting.

      {}

\section{CONCLUSIONS \& OUTLOOK}\label{sec:conclusions}
    In conclusion, model-free Reinforcement Learning is a promising approach towards data-driven optimal control for adaptive optics. We have shown how a closed-loop Recurrent Neural Network controller can be trained using the Recurrent Deterministic Policy Gradient algorithm, without any prior knowledge of the system dynamics and disturbances. This RNN only needs the most recent wavefront measurement and the previous DM commands at every time step. First, we applied the algorithm to tip-tilt control for a simulated, ideal AO system. We demonstrated that our approach can learn to mitigate a combination of tip-tilt vibrations, reducing the residual RMS by a factor of $\sim6$ as compared to an optimal-gain integrator. We also show a $\sim 25\%$ decrease in residual RMS for power-law input turbulence compared to the optimal gain integrator. These experiments were then repeated in a lab setup. For the vibration mitigation we observe a decrease in RMS by a factor 2.2 and a $\sim 25\%$ decrease in residual RMS for power-law input turbulence. Furthermore, we demonstrated that a single Recurrent Neural Network can identify and correct a vibration with varying amplitude and frequency without updating the control parameters.
    
    \subsection{Towards full wavefront control}
   In future work we will test the algorithm for the predictive control of a high-order DM. Our approach is almost directly applicable to full wavefront control, as we will only have to modify the reward function and the model architectures. To optimally make use of the spatio-temporal structure of the input data in full wavefront control, we propose to combine Recurrent Neural Networks and Convolutional Neural Networks (CNNs). As the reward for the full wavefront control, one could use a modal approach, where the modal residuals are separately minimized. In analogy to  our tip-tilt approach here, we can also locally minimize the spot deviation of a Shack-Hartmann wavefront sensor. Since we can use an arbitrary reward, it is, in principle, also possible to use a focal-plane quantity such as the Strehl ratio or contrast. However, the accurate estimation of these quantities at kHz frequencies might be an issue in practice.
    
    We showed that a single RNN can identify and mitigate vibrations with different frequencies and amplitudes without requiring an update of the control law. In a similar way, we expect it to be able to track changes in wind speed and direction when trained using data obtained during a variety of atmospheric conditions, as was shown in \citenum{Liu2020_LSTM}. This means that it would not require real-time updating of the control law. Since training of the controller can also be done completely separate from the data collection, this can, for example, even be done during the day. This implies that the computational demand of updating the controller is not an issue. Once a general model is learned, it would only require retraining once something in the instrument changes.
    
    In this work we started with a randomly initialized control law. However, we already empirically know what a reasonable control law will be. We could therefore already initialize these weights such that the output is equal to that of, for example, an integrator. Furthermore, we can also train on data collected using another controller. It is therefore possible to simply train on data collected using the current controller installed at the telescope. We can therefore already learn very reasonable models without even requiring time on the telescope, by learning from past data.
    
    While the computational demand is not an issue for tip-tilt control, it may become crucial for ELT-sized XAO systems. Even though the computation time is independent of the temporal order, the architecture used here would be expensive to apply to every mode. There is a trade-off between predictive power of the controller and the computational time. One could optimize the architecture given a constraint on the computation time for calculating the real-time commands.
    
    Finally, another advantage is the algorithm's ability to learn a nonlinear control law. It was recently shown that CNNs are able to reconstruct the nonlinearities of Fourier-based wavefront sensors \cite{Landman2020}. However, these reconstructors can only be calibrated on phase profiles spanned by the DM modes, while the response also depends on higher-order modes. Reinforcement Learning could allow us to calibrate a nonlinear controller on-sky. This may be useful for current and future instruments using a Pyramid Wavefront Sensor.

% References
\bibliographystyle{spiebib}
\bibliography{main} % bibliography data in report.bib
 % makes bibtex use spiebib.bst

\end{document}